\documentclass[aps,prl,onecolumn,superscriptaddress]{revtex4-2}
\usepackage[english]{babel}
\makeatletter
\let\lang@en\lang@english
\makeatother
\usepackage{graphicx}  
\usepackage{amsmath}
\usepackage{siunitx}
\usepackage{setspace} 
\usepackage{dcolumn}
\usepackage{bm}
\usepackage[mathlines]{lineno}

\usepackage[dvipsnames]{xcolor}

\usepackage{hyperref}              

\begin{document}
\title{Recovering the Shape of a Contact Line}

\thanks{E-mail: aka6449@psu.edu}%
\author{
Ashbell Abraham$^{1}$ Audrey Profeta$^{2}$ Jeanette Smit$^{2}$ Esmeralda Orozco$^{2}$ Charity Lizardo$^{2}$ Dani Medina$^{2}$ Aidan McGuckin$^{2}$ Bri Kroger$^{2}$ Shae Cole$^{2}$ Nathan Keim$^{1}$
\\
$^{1}$Department of Physics, The Pennsylvania State University, University Park, PA 16802, USA
\\
$^{2}$Department of Physics, California Polytechnic State University, San Luis Obispo, CA 93407, USA
\\
$^{3}$Commonwealth University-Lock Haven, Lock Haven, PA 17745, USA}



\begin{abstract}
We study the conditions for a three-phase contact line to return to a previous position. We drive a water-air-glass contact line between two horizontal plates, by slowly adding and removing water with a constant volume amplitude. For the first several cycles, the contact line ends each cycle with a different shape, in contrast with previously published work. Eventually the shapes begin to repeat, and the system has memory: a cycle with a smaller amplitude ends in a different shape, but even one cycle at the original amplitude recovers the steady-state shape. After a cycle at a larger amplitude, the steady-state shape is erased. We find that our tight control of the enclosed volume creates a global interaction, wherein only the least stable part of the contact line can move. Using theory and minimal models, we show that this interaction gives rise to the transient behaviors.  Our study sheds light on the origins of reversibility and memory in a system where neither is guaranteed, and shows that the physics of contact line motion changes in a confined environment. 
\end{abstract}



\maketitle

\twocolumngrid

\section*{Introduction}
The perimeter of a water drop sitting on a horizontal surface is typically rough and irregular, even as the drop's free surface is smooth and rounded. It is also stubborn: when the substrate is tilted slightly, or the volume of the drop is slightly increased or decreased, this liquid-solid-vapor contact line might not move at all. These behaviors are attributed to the microscopic roughness or heterogeneity of the substrate, whereby the contact line can be pinned at discrete locations scattered across its surface. Only when a segment of the contact line is depinned by a sufficient stress does it rapidly jump to another stable site. Despite this intuitive picture, modeling contact line physics is a longstanding challenge that requires attention to its details at nearly all length scales~\cite{yan_avalanches_2024,snoeijer_moving_2013}. Understanding contact line motion is especially important when surfaces must be coated, cleaned, or dried by dynamic wetting or dewetting. Recently, cycles of wetting and dewetting have received new attention, in part because of the importance of cyclic imbibition in porous media, in scenarios like depletion and replenishment of groundwater, cyclic waterflooding for enhanced oil recovery~\cite{li_using_2021}, and CO$_2$ sequestration~\cite{edlmann_cyclic_2019}. As fluid interfaces evolve, they may also redistribute adsorbed nutrients or pollutants in a system.

Cyclic motion would seem to amplify the intricacies of wetting and dewetting, as the contact line repeatedly navigates a vast, disordered landscape of possible pinned states. However, this scenario also prompts a simple question: will the contact line ever return to a previous position and shape? One answer comes from experiments and simulations by Holtzman et al., who showed that repeating a  cycle even once yields the same final shape, and that the motion is periodic in every subsequent repetition~\cite{holtzman_origin_2020,holtzman_relation_2023}. This straightforward behavior is known to arise from hysteresis and disorder---two prominent features of this system~\cite{holtzman_origin_2020,barker_magnetic_1997,sethna_hysteresis_1993,keim_memory_2019}. 
Similar ideas have long guided research on imbibition in porous media~\cite{haines_studies_1930}.
Notably, the experiments of Holtzman et al.\ are driven by varying the height of a connected reservoir, forcing the contact line up and down an inclined channel. Because of the reservoir's large free surface, its depth stays nearly constant even as the contact line moves in the narrow channel. Effectively, driving controls the pressure in the channel.

\begin{figure*}[tp]  
    \centering
        \includegraphics[width=\linewidth]{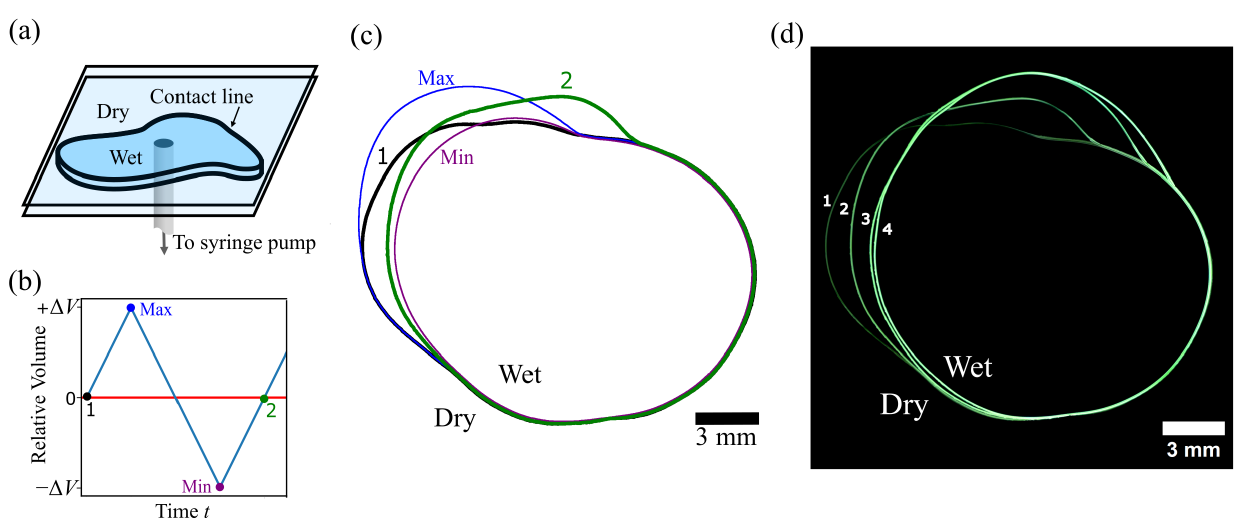}
        \caption{Non-periodic motion of a contact line in response to periodic driving. 
        \textbf{(a)} Deionized water is carried to and from the center of a glass channel via a syringe pump. 
        \textbf{(b)} The symmetric driving cycle illustrates how the relative volume reaches both turning points ($\pm \Delta V$) before returning to the initial volume. 
        \textbf{(c)} Contact line shapes at the labeled points in (b), with $\Delta V = 15$~$\mu$L. The connection to the syringe pump at the center is not shown.
        \textbf{(d)} Superposed photographs of 4 contact line shapes, each enclosing the same volume. Older shapes are darkened. Shapes 1 and 2 match (c); shapes 3 and 4 are after a second and third driving cycle.}
		  \label{fig:Figure1}
\end{figure*}

In this paper, we consider a system in which the only free surface is the narrow meniscus at the contact line. 
Rather than varying pressure, driving controls the enclosed volume tightly. 
We find that instead of immediately reaching a repeatable shape, the contact line takes on different shapes and trajectories for 10 or more cycles before it begins to repeat (Fig.~\ref{fig:Figure1}d). 
After this transient, the contact line exhibits return-point memory, encoding both the largest amplitude of driving and smaller amplitudes nested within it---a behavior that was always present in the pressure-controlled case~\cite{holtzman_origin_2020,holtzman_relation_2023}. By combining detailed observations 
with a simple kinematic model, we show that these behaviors are explained by volume conservation: when one section of the contact line advances, the pressure on other sections is relieved or even reversed.
In this case, the contact line does not belong to the large family of systems with return-point memory~\cite{keim_memory_2019}, but instead joins the more peculiar group that displays return-point memory after a transient, including glassy matter like amorphous solids~\cite{keim_generic_2011,keim_global_2020,keim_mechanical_2022}, crumpled sheets~\cite{matan_crumpling_2002,shohat_crumpled_2022}, and spin ice~\cite{gilbert_direct_2015}.
All these systems feature competitive or frustrated interactions among their elements.
These findings show that the apparent physics of a contact line depends on the mechanical environment, including all free surfaces. Our results also underscore the value of cyclic driving in revealing new aspects of contact lines and other matter that does not relax to equilibrium.

\begin{figure*}[tp]  
    \centering
        \includegraphics[width=\linewidth]{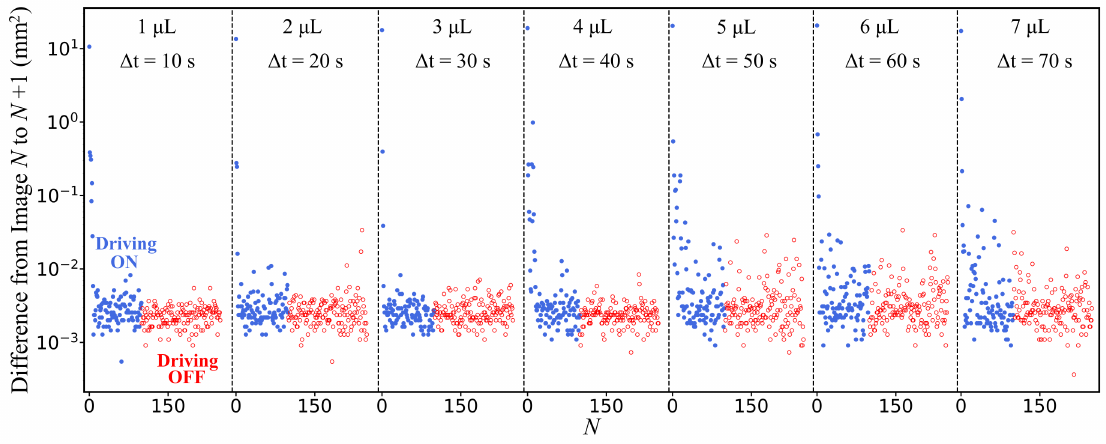}
        \caption{
			Evolution under cyclic driving. The contact line is driven for 100 cycles, followed by a period without driving, for 7 volume amplitudes. Labels at top are cycle amplitude $\Delta V$ and duration $\Delta t$. Blue closed points show image area changed by each driving cycle. Red open points also compare images taken $\Delta t$ apart, but with no driving, representing a baseline. A steady state is signaled by blue points becoming indistinguishable from red. At larger amplitudes, changes become small with intermittent fluctuations. 
			 }
		  \label{fig:Figure2}
\end{figure*}

\section*{Results}
\subsection*{Reaching a Steady State}

To encode a steady state, we ‘train’ the contact line by injecting and withdrawing a constant volume amplitude $\Delta V$ until it forms a periodic repeatable shape. This oscillatory driving is illustrated in Fig.~\ref{fig:Figure1}(b). As expected, the mobility of the entire contact line during an injection or withdrawal is not uniform; instead, we observe portions of the contact line, of arbitrary length, begin to slip while other regions remain pinned. This observable ‘stick and slip’ behavior is expected from a surface that is not microscopically smooth or homogeneous. At the macroscopic (averaged) scale, the liquid contact angle has an approximate range of stable values, and therefore the meniscus has a range of stable out-of-plane curvatures. Given that the mean curvature of the meniscus must be uniform, there is a corresponding approximate range of stable in-plane curvatures, as evidenced by the irregular shapes in Fig.~\ref{fig:Figure1}(c,d). Thus, while there is an effective line tension, its role is limited and the drop shape never becomes circular.

In Fig.~\ref{fig:Figure2}, we show how much the shape of the contact line changes after each cycle. Prior to reaching a steady state, there is an initial transient period during which the cyclic trajectories of the contact line will never repeat. However, as we apply more training cycles, the total number of pixels different eventually reaches a small value consistent with the subsequently measured noise, marking its transition into a steady state.

\subsection*{Memory Behavior}

\begin{figure}[h!]
\centering
\includegraphics[width=0.8\linewidth]{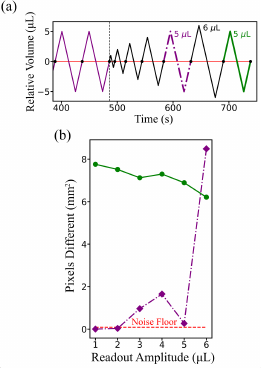}  
\caption{The steady state represents a memory that can be read and erased. 
\textbf{(a)} Protocol with the last 2 cycles of training at $\Delta V = \SI{5}{\micro\liter}$, followed by a ``ring up'' readout protocol of $\Delta V = 1$--\SI{6}{\micro\liter}, then a final cycle of \SI{5}{\micro\liter}.
\textbf{(b)} Memory signal (purple diamonds) is obtained by comparing the image before readout with the image after each readout cycle, plotted as a function of amplitude. The red dashed line represents the steady-state noise floor. Green circles compare each readout image to the image after the final \SI{5}{\micro\liter} cycle, showing that the memory of training is erased.}

\label{fig:Figure3}
\end{figure}

At the end of the transient, our system reaches a steady state in which further cycles of driving leave its shape unchanged. We now test the properties of this steady state with the protocol in Fig.~\ref{fig:Figure3}(a). After training for 100 cycles with amplitude 5~$\mu$L, we reduce the amplitude to 1~$\mu$L and then apply cycles of increasing amplitude, up to 6~$\mu$L. After each cycle, we compare the shape of the contact line with the shape at the end of training. We plot the difference as a function of volume amplitude in the ``Readout'' curve of Fig.~\ref{fig:Figure3}(b). Applying smaller-amplitude cycles disrupts the steady state, leaving the contact line in a different shape even as it returns to the same volume. The maximum difference in Fig.~\ref{fig:Figure3}(b), after applying a cycle of 4~$\mu$L, is order 1~mm$^2$, comparable to the large differences early in each transient in Fig.~\ref{fig:Figure2}. However, when we apply just one cycle of the original 5~$\mu$L training amplitude, the contact line recovers its steady-state shape, while increasing the amplitude further to 6~$\mu$L produces a sharp difference. This result shows that the steady state encoded a memory of the training amplitude that formed it, which is indicated by the local minimum in Fig.~\ref{fig:Figure3}(b).

The readout behavior in Fig.~\ref{fig:Figure3} is consistent with return-point memory, a property of many systems that contain hysteretic elements~\cite{keim_memory_2019}. This property means that while driving is bounded by any two turning points (here, $\pm 5$~$\mu$L), revisiting either turning point must restore the system to the state it previously had at that turning point. Thus when the readout amplitude is increased to 5~$\mu$L, the system resumes the trajectory it had at the end of training, and returns to the same state at the end of the cycle. If the bounds are exceeded, however (here, by the 6~$\mu$L cycle), return-point memory will generally \emph{not} hold. We verify this by applying one final cycle of 5~$\mu$L. The resulting shape is markedly different from every shape before it (upper curve in Fig.~\ref{fig:Figure3}a), showing that the original memory is erased and the previous steady state is inaccessible.

\subsection*{Effects of Volume Control}
\begin{figure*}[tp] 
    \centering
        \includegraphics[width=\linewidth]{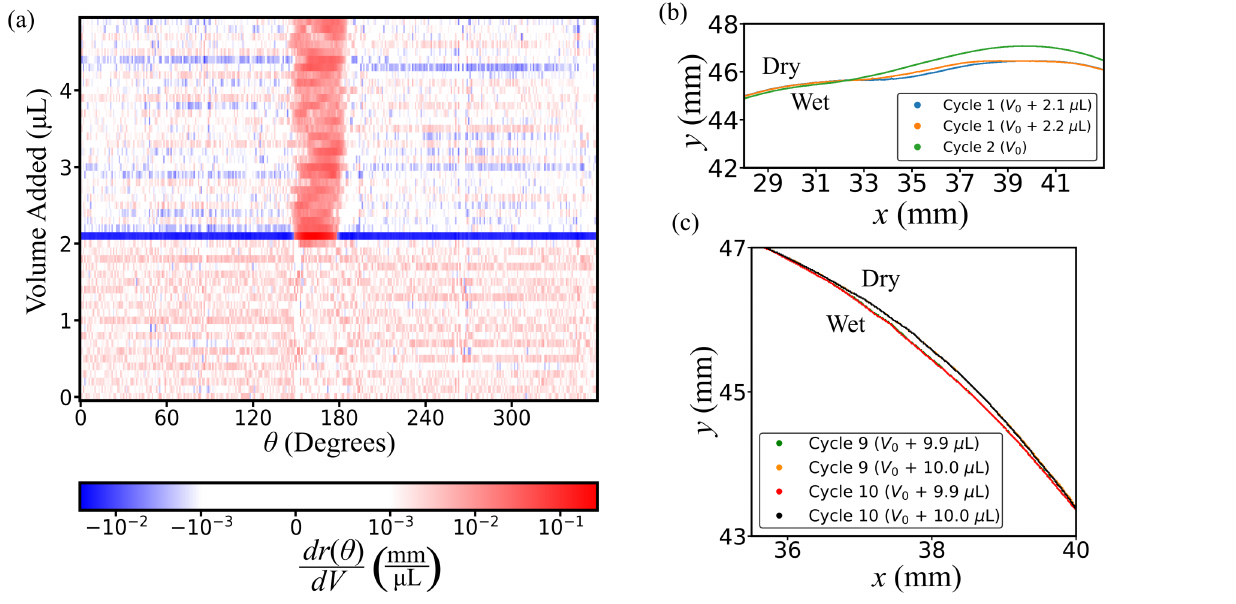}
        \caption{
    		Volume conservation. 
            \textbf{(a)}  
            We inject water while capturing an image every \SI{0.1}{\micro\liter}. 
            The displacement of each $1^\circ$ 
            segment of the contact line is shown for the first 50 images, starting at the bottom of the plot. An abrupt local advance (red) at $V_0 + \SI{2.2}{\micro\liter}$ is accompanied by a pressure drop that appears to pull back the rest of the contact line (blue). Values below our resolution ($\sim$ $10^{-3}$ mm) are mapped to white. Vertical streaks (e.g.\ near $250^\circ$) are from small-scale image flaws; the contact line remains smooth at these locations. 
            \textbf{(b)} A portion of the contact line is shown before and after the large jump in (a) (``Cycle 1''). ``Cycle 2'' is the shape upon further injecting to $V_0 + \SI{20}{\micro\liter}$ and withdrawing back to the original $V_0$. Although at a lower volume, the ``Cycle 2'' contact line is positioned well above the location of the earlier large jump---the event will not repeat.
            \textbf{(c)} An example of a large jump that repeats at the same volume in two consecutive cycles, in the steady state reached with protocol $V_0 \rightarrow V_0 + {\SI{10} {\micro\liter}} \rightarrow V_{0}$.}
		  \label{fig:Figure4}
\end{figure*}

To shed light on the process of reaching a steady state, we examine the unsteady ``stick-slip'' motion of the contact line as it is trained. The syringe pump injects and then withdraws \SI{20}{\micro\liter}, while triggering the camera for every \SI{0.1}{\micro\liter}. We track the shape of the contact line  in radial coordinates, $r(\theta)$, measured every 1$^\circ$, with the origin fixed at the initial centroid of the water. Fig.~\ref{fig:Figure4}(a) shows the differential change $dr(\theta) / dV$ due to each volume step, with a color scale that emphasizes decreases in $r$. Initially, injection causes a slight apparent outward motion of the contact line. However, upon reaching 2.2~$\mu$L, a small region advances roughly 100 times more, and the rest of the contact line appears to recede. The advancing segment is fed by volume from the rest of the free surface: other segments either depin and recede, or---more likely for the observed uniform motion---the meniscus contracts due to the abrupt drop in Laplace pressure~\cite{rost_fluctuations_2007}. Several subsequent injection steps repeat this pattern with smaller magnitude.

The most prominent depinning events, like the one in Fig.~\ref{fig:Figure4}(a), typically occur early in training and do not repeat in subsequent cycles. In Fig.~\ref{fig:Figure4}(b) we plot the contact line at the site where it advances, before and after this jump at 2.2~$\mu$L in Fig.~\ref{fig:Figure4}(a). At the same site on the next cycle, part of the shape for a \emph{smaller} enclosed volume lies \emph{outside} these contours, showing emphatically that the trajectory is not yet periodic. Once the macroscopic motion of the contact line becomes roughly periodic, we can identify large and repeatable depinning events. An example of a repeating jump is in Fig.~\ref{fig:Figure4}(c), where the large number of pixels that overlap shows that, once trained, the contact line repeats its past motion precisely.

\section*{Model}

To understand the qualitative results from our experiments, we propose a simple model of contact line hysteresis that we simulate in the volume-controlled and pressure-controlled limit. We treat the contact line as an ensemble of $N$ hysteretic elements, each of which models the motion of a  segment. The position of the $k$th segment is the integer coordinate $j_k$, which we increment or decrement by 1 to represent the advancing or receding of that segment. 
For convenience we represent our system on a 2D grid (Fig.~\ref{fig:Figure5}a) with axes $k$ and $j$; however $k$ is purely an index and does not correspond to physical locations. Our simplified model does not include line tension, by which each element would be coupled to its nearest neighbors. This would give rise to avalanches in which the motion of one element destabilized others; instead, our model coarse-grains avalanches at the level of the discrete elements. 
More formally, such cooperative or ``ferromagnetic'' interactions are known to be fully compatible with return-point memory, and they cannot cause extended transients~\cite{sethna_hysteresis_1993,holtzman_origin_2020} (details in Supporting Information).

The ease or difficulty of changing a $j_k$ is represented by two thresholds at each cell on the grid. Similar to advancing and receding contact angles in the physical system, elements at cells with lower advancing thresholds $A_{jk}$ advance most readily, while elements at cells with \emph{higher} receding thresholds $R_{jk}$ are first to recede. To represent the substrate disorder, $A_{jk}$ and $R_{jk}$ are from a normal distribution with standard deviation $\sigma=100$. In all simulations conducted, the total number of elements is $N=100$.

\subsection*{Pressure-Controlled Simulation} 
For pressure-controlled simulations, we emulate an inclined channel by adding an effective hydrostatic pressure term $gj$ to the thresholds~\cite{holtzman_origin_2020}. With $g > 0$, elements at higher $j$ tend to be harder to advance and easier to move back. We use $g = 1/4$, which allows disorder to dominate while preventing runaway depinning events in which elements could leave the simulation domain. We initialize the elements at $j_k = 0$ and advance them until they are stable at an initial pressure $P_0$, at positions roughly halfway along the $j$ axis.

In each cycle we vary the pressure $P$ between $P_\text{min} = P_0 - \Delta P$ and $P_\text{max} = P_0 + \Delta P$, in the sequence $P_0$ $\to P_\text{max}$ $\to P_\text{min}$ $\to P_0$, where $\Delta P = 100$. As we increase $P$ we advance each element independently, as long as its $A_{jk} \leq P$; for decreasing $P$ we recede while $R_{jk} \geq P$. In Fig.~\ref{fig:Figure5}(b) we show that the transient length $\tau$, defined as the number of cycles needed to reach a steady state, is at most 1 cycle long in our pressure-controlled model---a result we obtain reliably for various $\sigma$ and $g$. A ring-up protocol analogous to that of Fig.~\ref{fig:Figure3}(a) lets us read out $\Delta P$ in Fig.~\ref{fig:Figure5}(c); the curve reaches zero at $\Delta P$,  as expected from return-point memory~\cite{barker_magnetic_1997,keim_global_2020}. The exact shape of the curve depends on $\sigma$ and $g$. In the limit of strong gravity, $g \gg \sigma$, the system no longer has hysteresis and the steady state will have uniform $j_k$; driving past the training amplitude does not erase the memory. In the limit of strong pinning, $g \ll \sigma$, the readout curve remains at zero until it is driven past the training amplitude, and then becomes large.

The hysteretic elements in our pressure-controlled model do not interact with each other; therefore, we may explain the one-cycle transient by considering the possible trajectories of a single element. Before the first driving cycle, the element advances from $j_k=0$ to an initial position that is stable at $P_0$. Afterwards, the pressure is increased to $P_{max}$ and then subsequently lowered to $P_{min}$. At this point, $\tau =1$ is guaranteed if the element is above its initial position, and raising the field back to $P_0$ will not advance the element any further. In contrast, if the element is below its initial position, raising the field from $P_{min} \to P_0$ advances the element back to its initial position, in which case $\tau = 0$; however, this is a special case that we do not observe in our simulations, as there is always at least 1 element that ends above its initial position at $P_{min}$. Through a different analysis, Holtzman et al.~\cite{holtzman_origin_2020} showed that in their pressure-controlled experiments and simulations, a repeating steady state was also achieved after the first driving cycle and return-point memory would emerge as a consequence. In the Supporting Information we further relate our analysis to formal treatments of return-point memory and hysteresis in other systems~\cite{sethna_hysteresis_1993,middleton_asymptotic_1992}.

\subsection*{Volume-Controlled Simulation}
\begin{figure*}[tp]  
    \centering
        \includegraphics[width=\linewidth]{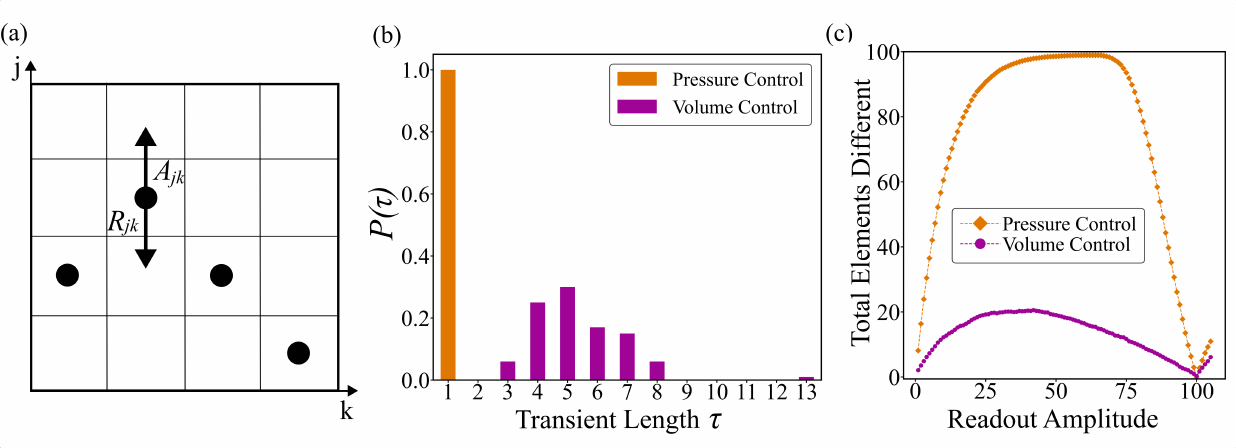}
        \caption{Contact line simulation.
        \textbf{(a)} Illustration of a 2-D lattice where each cell is given a unique advancing and receding threshold.
        \textbf{(b)} Transient lengths for a system of $N=100$ elements and their probability of occurring in 100 simulations, for both pressure and volume control.
        \textbf{(c)} The number of elements at positions different from the start of readout, as a function of readout amplitude, analogous to Fig.~\ref{fig:Figure3}b, showing recovery of the training amplitude. Each curve is the average of 100 runs. 
        }
		\label{fig:Figure5}
\end{figure*}

In the volume-controlled limit, the elements begin halfway along the $j$ axis with uniform $j_k = 2500$. The total enclosed volume is $\sum_k j_k$, which driving changes in discrete steps: each advancing step involves finding the element with the smallest present value of $A_{jk}$, and incrementing its $j_k$ by 1; receding decrements $j_k$ at the element with the largest $R_{jk}$. The disorder $\sigma$ must be nonzero to break degeneracy, but its value is irrelevant. A complete cycle of driving consists of $\Delta V$ advancing steps, $2\Delta V$ receding steps, and $\Delta V$ advancing steps where $\Delta V = 100$ steps. In Fig.~\ref{fig:Figure5}(b), we show that $\tau$ takes on a range of values greater than 1, inconsistent with return-point memory. Tests with $N=100$ and 1000 show that $\tau$ depends only weakly on system size. After reaching a steady state, we can reliably read out $\Delta V$ using a ring-up protocol like in Fig.~\ref{fig:Figure3}(a). 
When we average many simulation readouts, the portion below $\Delta V$ shown by Fig.~\ref{fig:Figure5}(c) resembles the downward-concave curve of the pressure-controlled case, except the minimum at $\Delta V$ is approximately 0.3 elements out of 100. Thus the memory-encoding state \emph{cannot} always be recovered exactly.

Controlling volume, rather than pressure, is equivalent to a global (i.e.\ mean-field), anti-ferromagnetic interaction among elements: when one element advances or recedes, it inhibits all others from doing so~\cite{muhaxheri_bifurcations_2024}. Anti-ferromagnetic interactions can lead to violations of return-point memory, of which extended transients and inexact readouts are examples. By contrast, the line tension in the physical system is a ferromagnetic interaction, in which one element's transition facilitates others', and it does not degrade return-point memory. 
The roles of interactions were first analyzed by Sethna et al.\ in the random-field Ising model~\cite{sethna_hysteresis_1993}, and in the Supporting Information we apply a similar analysis to our model. Briefly, return-point memory requires ``no-passing''~\cite{middleton_asymptotic_1992}: if the contact line shape $r$ is contained completely within shape $s$, applying cycles of driving to each should not change their ordering. Here, a portion of $r$ can overtake $s$ if another portion falls behind. 

\begin{figure}[h!]
\centering
\includegraphics[width=0.5\linewidth]{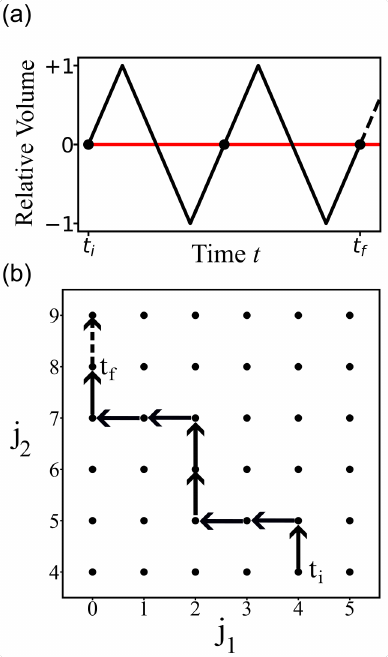}  
\caption{Considering a pair of elements shows how anti-ferromagnetic coupling allows extended transients.  
\textbf{(a)} A notional cyclic driving protocol like those in experiments, with amplitude 1.
\textbf{(b)} The positions of elements 1 and 2 are represented on axes $j_1$ and $j_2$. When following the driving protocol from (a) on a lattice with thresholds that obey Eq.(\ref{eq:Threshold_Inequality}), the elements follow an open trajectory that will never reach a steady state.}
\label{fig:Figure6}
\end{figure}

To illustrate the role of the antiferromagnetic interaction, we construct a two-element ``ratchet'': when driven by the cyclic protocol in Fig.~\ref{fig:Figure6}(a), it will end each cycle in a new state. The smallest advancing thresholds and largest receding thresholds belong to elements 2 and 1 respectively,
\begin{equation}
\begin{aligned}
\min_j \left( A_{j1} \right) &> \max_j \left( A_{j2} \right),\\ 
\min_j \left( R_{j1} \right) &> \max_j \left( R_{j2} \right).
\end{aligned}
\label{eq:Threshold_Inequality}
\end{equation}
This construction leads to arbitrarily large $\tau$, represented in Fig.~\ref{fig:Figure6}(b). Because only the least stable element can move, element 1 never advances and element 2 never recedes. 
Extended transients can thus arise from regions matching Eq.~\ref{eq:Threshold_Inequality} that occur by chance within a larger domain. Beyond this minimal example, there may be many more ways that extended transients arise, even in small systems---just as previous studies have found for elements that transition between only two states~\cite{lindeman_glassy_2021,keim2021,van_profusion_2021}. Nonetheless, we occasionally observe ratchet-like behavior in experiments, particularly at small initial volumes ($< 0.5$~mL): the drop's center of mass migrates in a single direction, and the transient seems indefinite. 

\section*{Conclusion}

We have shown that by isolating our liquid contact line system so that its volume is controlled tightly, we change the global physics of its motion. Many cycles of driving are required for the contact line's shape to become repeatable. This result would seem to contradict a more conventional picture of contact line motion involving hysteresis, quenched disorder, and cooperative interactions (i.e.\ line tension). These generic ingredients are known to preclude transients and give a system return-point memory~\cite{sethna_hysteresis_1993}, as confirmed by the simulations and experiments of Holtzman et al.\ that drove a contact line by varying hydrostatic pressure~\cite{holtzman_origin_2020,holtzman_relation_2023}. 
 
We traced the transient we observe to an additional essential feature of contact line motion under controlled volume: as one local segment of the contact line jumps forward, the outward pressure on other regions is reduced or even reversed. The volume constraint thus acts as a global antiferromagnetic, frustrated, or competitive interaction among segments of the contact line. Because of this interaction, return-point memory is not guaranteed~\cite{sethna_hysteresis_1993,terzi_state_2020,bense_complex_2021}, and transients are possible.  Our experiments and simulations show that transients are indeed common, but that the steady-state behavior nonetheless shows return-point memory. The simulations also show that these behaviors do not require cooperative or ferromagnetic interactions between neighboring segments, so that we may neglect line tension. However, in experiments these local interactions are responsible for avalanches, during which we observe the \emph{global} interaction most strongly (Fig.~\ref{fig:Figure4})---suggesting that local cooperative interactions are not in conflict with global competition, and in fact may amplify its effects.

Our results suggest that the mechanical compliance of the environment can be important in complex multiphase flows---including flows of immiscible fluids within porous media in groundwater hydrology, oil recovery, and carbon dioxide sequestration. A low compliance, $dV/dP$, means that local changes in volume are accompanied by large pressure changes~\cite{sun_haines_2019}. 
Our model suggests that one should consider whether the depinning of one segment changes the pressure enough to suppress further depinning---for example, a characteristic volume for a jump, $\Delta V$, could be compared with $\Delta P dV/dP$, where $\Delta P$ is roughly the pressure difference between advancing and receding in a pressure-controlled experiment.
We note that in porous media, the effective $dV/dP$ can depend on timescale: compliance is increased by the presence of free surfaces elsewhere, but is lowered if, for example, those surfaces are connected via a constricted channel, inhibiting rapid volume changes. 
Thus it could be that even as pressure is varied cyclically at the boundary of a porous medium, at the mesoscopic scale, contact line segments compete for volume.
As a further generalization, while we have focused on periodic flows or deliberate amplitude variation, results from disordered solids suggest that even when driving fluctuates randomly, our system may also evolve toward a steady state and form memories~\cite{mungan_random_2025}. A further open question in porous media is whether the transient evolution is accompanied by a change in dissipation or relative permeability~\cite{keim_memory_2019,paulsen_mechanical_2025,shohat_dissipation_2023}.

The volume-controlled contact line joins other systems with apparent return-point memory after a transient~\cite{paulsen_mechanical_2025}, including artificial magnetic systems~\cite{gilbert_direct_2015}, amorphous solids~\cite{keim_generic_2011,keim_global_2020,keim_mechanical_2022,khirallah_model_2021}, crumpled sheets~\cite{matan_crumpling_2002,shohat_crumpled_2022}, and small groups of interacting bistable elements~\cite{lindeman_glassy_2021,paulsen_tunable_2026}. Each of these examples features frustrated or antiferromagnetic interactions that are pairwise and local, but here a global interaction is key. Our system thus resembles experiments that control the length of a chain of fabricated bistable elements~\cite{bertoldi_flexible_2017,kwakernaak_counting_2023}, or models that prescribe the total volume of several bistable balloons~\cite{muhaxheri_bifurcations_2024}. However in our study, rather than specifying the properties of a small number of bistable elements, we obtain a memory behavior from the naturally-occurring microscopic disorder of the substrate; each phase in our system appears featureless, and states are bounded only by the size of the channel. We may ask whether other enhanced memory and computational behaviors enabled by frustrated interactions~\cite{lindeman_generalizing_2025, keim2021,lindeman_glassy_2021, kwakernaak_counting_2023, paulsen_mechanical_2025} are also naturally present in this distinct type of system. 
Finally, our work suggests that just as tunable mechanical metamaterials~\cite{liu_pathways_2024,paulsen_tunable_2026} and flow networks~\cite{altman_flow_2026} offer a variety of tailored behaviors rooted in hysteresis, interacting menisci in fabricated channels ~\cite{moebius_interfacial_2012,sun_haines_2019}  could become a coequal platform for memory and computation. 


 \section*{Methods}
 In the experimental setup of Fig.~\ref{fig:Figure1}(a), we capture images of the contact line's shape inside the gap ($h_\text{gap}$= 0.75~mm) of a Hele-Shaw cell made with soda-lime glass plates, each with dimensions of 140~mm $\times$ 140~mm $\times$ 6.35~mm. Although there are two contact lines in this system, one on each plate, we assume that both have the same shape. The bottom plate has a 6.35~mm diameter hole drilled at its center, into which we tightly fit a PTFE tube with slightly larger diameter. The PTFE tube is connected with a brass compression fitting to a semi-rigid nylon water line, which leads to a 1~mL gas-tight syringe on a Cetoni Nemesys syringe pump that has $\leq 40$~nL repeatability (verified independently). Before each experiment, both plates are cleaned using acetone (Fisher Scientific, HPLC Grade, cat.\ no.\ A9494), ethanol (Fisher Scientific, ACS/EP/USP Grade, cat.\ no.\ A4094), deionized water, and 2-propanol (Fisher Scientific, HPLC Grade, cat.\ no.\ A4514) and then immediately dried with nitrogen ($< 2$~ppm moisture). Cleaning with 2-propanol has been reported to increase contact line roughness; however, this behavior is a mild effect and does not impede the goal of our experiment~\cite{schutt_capillary_2001}. Four damp pieces of paper towel are placed near the edges to control evaporation. After injecting water to the surface of the bottom plate, $\sim$1.0~mL of water is dispensed on top of the PTFE tube with a handheld syringe. Afterwards, we seal the edges of the cell with weatherproof sealing tape (McMaster-Carr, cat.\ no.\ 1228a56) and clamp each corner. Images of the contact line are captured by a Nikon D5300 camera with a lens of focal length 100~mm, mounted above the cell. The experiment is illuminated by a flexible LED strip wrapped around the edges of the cell.

The contact line is cyclically driven by injecting and withdrawing a constant ``training'' volume ($\Delta V$), from the initial water drop placed within the cell. The driving flow rate is kept at 0.416~$\mu$L/s, keeping viscous stresses negligible compared to capillary stresses, as measured by the capillary number: $\mathrm{Ca} = \frac{\mu U}{\sigma} < 10^{-8}$, where $\sigma$ and $\mu$  are the surface tension and dynamic viscosity of water at room temperature, respectively, and $U$ is the characteristic velocity of the contact line. After each experimental trial, the starting volume is changed by injecting a volume that is of the same order of magnitude as the previous training volume. 

Our camera is set to capture images of the contact line at the end of every driving cycle, after a 1.0~s pause to ensure that the system has come to rest. We compare two images by subtracting each pixel’s grayscale value; if this difference exceeds a threshold, we consider those two pixels to be different. We choose the lowest threshold that does not result in a spurious signal from the static region at the center of the image; larger thresholds yield smaller but qualitatively consistent signals. We express the image difference as an area in mm$^{2}$, where one pixel is $1.8 \times 10^{-4}$~mm$^2$. We estimate the combined noise from our camera, illumination, external vibrations, and evaporation by measuring how much the contact line changes over time $\Delta t$ without driving. The values of $\Delta t$ range from 10 to 70~s, to match the duration of a driving cycle over the range of volume amplitudes. By performing this baseline measurement immediately after each cyclic driving test, we also account for slow variation in the evaporation rate and vibrational noise levels. If the counts of changed pixels near the end of driving are consistent with our noise measurements, we conclude that the contact line has reached a periodic steady state.

\section*{Data Availability}
All data used in this study, including raw and analyzed images, Python codes, and tabular data (CSV and TXT files) have been deposited in Zenodo \\(https://doi.org/10.5281/zenodo.19653337)

\section*{Acknowledgements}
We thank Omri Barak, Aaron Bock, Karin Dahmen, Sujit Datta, Anthony Dinsmore, Alexander Johnson, Angel Martinez, Muhittin Mungan, Nate Martin, and Ling-Nan Zou for helpful discussions and technical assistance. This work was supported by NSF DMR-1708870 (A.P., C.L., A.M., D.M., B.K., E.O., and N.K.), DMR-2011839 and DMR-1851987 (S.C), and by the Human Frontier Science Program grant RGP0017/2021 (A.A.).

\bibliography{PRX_Submission/UpdatedCitations3_30_2026}
\clearpage
\onecolumngrid   

\begin{center}
    \LARGE \textbf{Supporting Information for\\[6pt]
    ``Recovering the Shape of a Contact Line''}
\end{center}
\vspace{1em}
\clearpage

\appendix
\setcounter{section}{0}
\renewcommand{\thesection}{S\arabic{section}}
\refstepcounter{section}   

\section{Syringe Pump Precision}

\setcounter{figure}{0}
\renewcommand{\thefigure}{S\arabic{figure}}
\begin{figure}
\centering
\includegraphics[width=\textwidth]{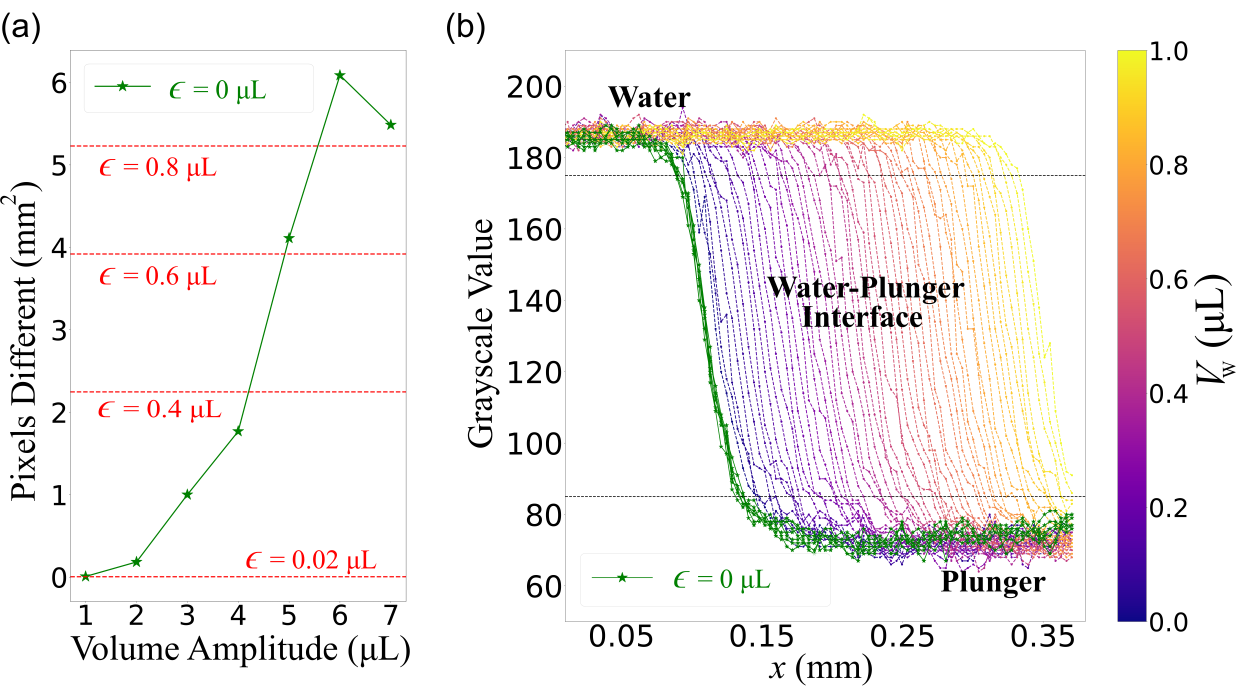}
\caption{Contact line shape changing signal independent of syringe error. \textbf{(a)} The green stars show how much the contact line changes once it returns back to its initial volume after being driven (symmetrically) at the volume amplitude. The red dashed line shows how different two images of the contact line are expected to be when separated by a volume $\epsilon > 0 $ $\mu L$.
\textbf{(b)} For a separate cyclic driving experiment (for amplitudes 1--\SI{7}{\micro\liter}), the green stars show how the plunger's position changes after returning to the initial volume at the end of each cycle. The other curves (color bar at right) show the plunger position during the constant withdrawal experiment, where the color of each curve indicates the total volume withdrawn $V_w$. Neighboring curves are separated by $ 0.02$ $\mu L$.}
\label{fig:SupplementaryFigure1}
\end{figure}

The syringe pump is not a significant source of systematic error or hysteresis in the observed contact line motion. To show this, we suppose that at the end of each cycle in our cyclic driving experiments, the contact line will return to its initial volume plus some mechanical error from our syringe pump, $V_i + \epsilon $. This error would also be apparent in the position of the syringe plunger. Thus, to show whether the contact line behavior we report could be explained by syringe pump error, we will estimate the $\epsilon$ that could cause the observed changes in the contact line shape (e.g.\ Fig.~2 in main text) and compare it with a direct measurement of $\epsilon$ from the position of the syringe plunger. In Fig.~\ref{fig:SupplementaryFigure1}a we show how much the contact line changes after each driving cycle where the volume amplitude increases in steps of 1~$\mu$L, similar to the first data point at each amplitude in Fig.~2 of the main text. We also plot, as horizontal lines, the effect of deliberately withdrawing a small volume, emulating the effect of a nonzero $\epsilon$ at the end of a cycle. These results show that a very large volume error, $\epsilon \sim 1$~$\mu$L, is required to create signals comparable to the transient at the start of cyclic driving. 

During the same cyclic driving and constant withdrawal protocols, we use the camera and macro lens to track the location of the syringe plunger. From this direct measurement we find that $\epsilon \ll 1$~$\mu$L. In Fig.~\ref{fig:SupplementaryFigure1}b we sample the image intensity along the axis of the syringe; the drop from light to dark corresponds to the plunger-water interface. The green squares show how the location of the syringe plunger typically varies in the cyclic driving experiments. The dashed lines (see color scale) show how the location of the syringe plunger changes throughout the constant withdrawal experiment. For the cyclic driving experiments, the plunger returned to its initial location with $<0.04$ $\mu $L precision---roughly half the smallest volume step used for the experiments in our main text. These results confirm that mechanical error from our syringe pump is not the source of the history-dependent behaviors we report.

\refstepcounter{section}   
\section{Conditions for Return-Point Memory}
\subsection{Pressure Control Forbids Extended Transients}
We expect that the transient length $\tau$ is at most 1 cycle, because our pressure-controlled model satisfies the same conditions as the proof of return-point memory in the athermal random-field Ising model by Sethna et al.~\cite{sethna_hysteresis_1993}. One of these conditions requires this system to evolve quasi-statically, which is implied by the rules in our course-grained kinematic model and can be achieved experimentally by slowly varying the pressure (negligible viscous dissipation). With this established, we may further show that our model satisfies Middleton's ``no-passing'' rule~\cite{middleton_asymptotic_1992}, which is another crucial requirement for return-point memory: If the pressure field $P_s \ge P_r$, then the set of element positions in $s\ge r$, given that $s$ and $r$ obtained by the fields $P_s$ and $P_r$, respectively. This property forbids an element in $s$ to be found in a position lower than it was at $r$. No-passing is guaranteed by the stability condition implemented in our pressure-controlled model: that advancing (receding) a stable element always requires an increase (decrease) in pressure. Once the pressure field is increased (decreased) to $P$, the elements may come to rest---independently from each other--- after reaching a threshold that is greater (less) than or equal to $P$. 

In the physical system, line tension creates a ferromagnetic coupling between neighboring contact line segments. This cooperative interaction means that the stability of any segment will depend on the stability of its neighbors and is a feature we do not include in either model. However, the dynamics introduced  by this interaction does not disrupt the partial ordering of states and maintains the no-passing property, a result shown for the random-field Ising model by Sethna et al.~\cite{sethna_hysteresis_1993} and for the contact line by Holtzman et al.~\cite{holtzman_origin_2020}.

\subsection{Volume Control Allows Extended Transients}

\begin{figure}
\centering
\includegraphics[width=\textwidth]{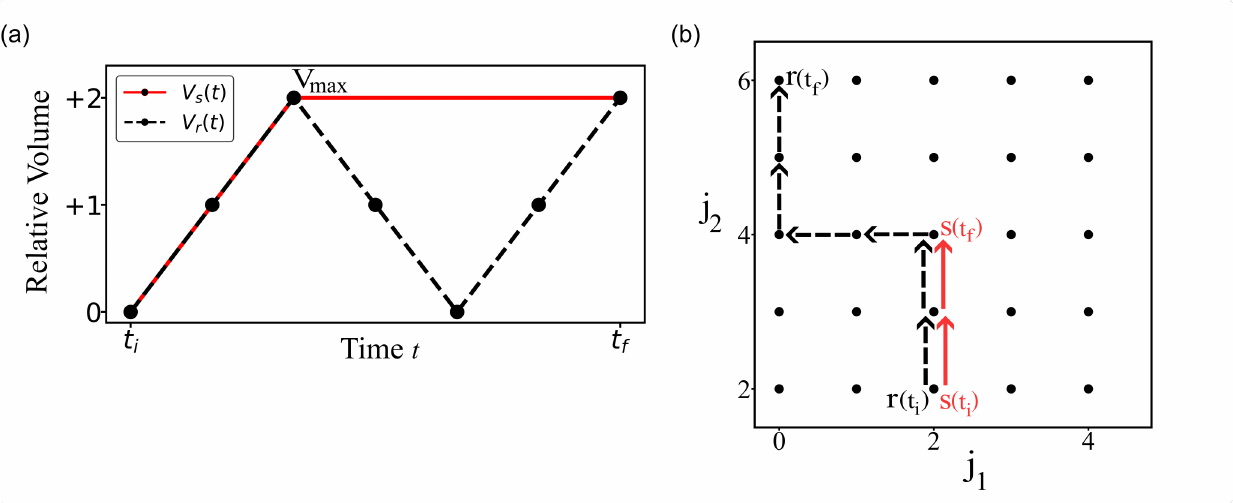}
\caption{Consistent with extended transients, the two-element ratchet violates Middleton's ``no-passing'' condition~\cite{middleton_asymptotic_1992}. 
\textbf{(a)} Monotonic and non-monotonic protocols $V_s(t)$ and $V_r(t)$ go from 0 to $V_{\max}$, where $V_{\max}=2$. 
\textbf{(b)} The positions of elements 1 and 2 are represented on axes $j_1$ and $j_2$. The elements' thresholds obey Eq.~\ref{eq:Threshold_Inequality}. When driven with protocols $V_r(t)$ and $V_s(t)$, the elements follow trajectories $r(t)$ and $s(t)$. The final positions are different even with the same initial condition $(j_1,j_2)=(2,2)$. $V_r(t)$ never exceeds $V_s(t)$ and ends at the same value, yet $j_2$ in $r(t_f)$ surpasses $j_2$ in $s(t_f)$, violating Middleton's no-passing rule~\cite{middleton_asymptotic_1992} and permitting violations of return-point memory~\cite{sethna_hysteresis_1993}, including extended transients.}
\label{fig:SupplementaryFigure2}
\end{figure}

We may further use our two-element ratchet to show that our volume-controlled model does \emph{not} allow a proof of return-point memory similar to that of Sethna et al.~\cite{sethna_hysteresis_1993}. Here, the smallest advancing thresholds and largest receding thresholds belong to elements 2 and 1 respectively,

\begin{equation}
\begin{aligned}
\min_j \left( A_{j1} \right) &> \max_j \left( A_{j2} \right),\\ 
\min_j \left( R_{j1} \right) &> \max_j \left( R_{j2} \right).
\end{aligned}
\label{eq:Threshold_Inequality}
\end{equation}
Figure~\ref{fig:SupplementaryFigure2}a shows two driving protocols $V_r(t)$ and $V_s(t)$ with the same initial volume, final volume, and bounds.
Nonetheless, the final state of the system depends on which driving protocol is used, where $V_s(t)$ and $V_r(t)$ result in a final state $s(t_f)$ and $r(t_f)$, respectively. We show in Fig.~\ref{fig:SupplementaryFigure2}b that, when $V_r(t)$ returns to $V_{\max}$, element 2 advances past its position at $s(t_f)$---a direct violation of Middleton's no-passing rule, made possible because element 1 receded. This violation means that states may not be partially ordered: given $s(t_i) = r(t_i)$ and $V_s(t) \ge V_r(t)$, it does not follow that $s(t_f) \ge r(t_f)$ for all elements. Thus return-point memory cannot be guaranteed~\cite{sethna_hysteresis_1993,mungan_structure_2019}.


\end{document}